\documentclass[a4paper,12pt]{article}

\newcommand{\sect}[1]{\setcounter{equation}{0}\section{#1}}

\newcommand{\bea}{\begin{eqnarray}}
\newcommand{\eea}{\end{eqnarray}}
\newcommand{\be}{\begin{equation}}
\newcommand{\ee}{\end{equation}}
\newcommand{\vs}[1]{\vspace{#1 mm}}

\newcommand{\dsl}{\pa \kern-0.5em /}

\newcommand{\pa}{\partial}

\newcommand{\nn}{\nonumber\\}

\newcommand{\tA}{\tilde{A}}
\newcommand{\tB}{\tilde{B}}
\newcommand{\tC}{\tilde{C}}
\newcommand{\tD}{\tilde{D}}
\newcommand{\tg}{\tilde{g}}
\newcommand{\tf}{\tilde{f}}
\newcommand{\tc}{\tilde{c}}
\newcommand{\talpha}{\tilde{\alpha}}
\newcommand{\tbeta}{\tilde{\beta}}
\newcommand{\hA}{\hat{A}}
\newcommand{\hB}{\hat{B}}
\newcommand{\hC}{\hat{C}}
\newcommand{\hD}{\hat{D}}
\newcommand{\hg}{\hat{g}}
\newcommand{\hf}{\hat{f}}
\newcommand{\hc}{\hat{c}}
\newcommand{\halpha}{\hat{\alpha}}
\newcommand{\hbeta}{\hat{\beta}}
\begin{document}
\topmargin 0pt
\oddsidemargin 0mm

\begin{flushright}
hep-th/0305175\\
\end{flushright}

\vs{2}
\begin{center}
{\Large \bf  
Dimensional reductions of M-theory S-branes to string theory S-branes}
\vs{10}

{\large Shibaji Roy\footnote{E-Mail: roy@theory.saha.ernet.in}}
\vspace{5mm}

{\em 
 Saha Institute of Nuclear Physics\\
 1/AF Bidhannagar, Calcutta-700 064, India\\}
\end{center}

\vs{10}
\centerline{{\bf{Abstract}}}
\vs{5}
\begin{small}
We study both the direct and the double dimensional reduction of space-like
branes of M-theory and point out some peculiarities in the process unlike
their time-like counterpart. In particular, we show how starting from SM2
and SM5-brane solutions we can obtain SD2 and SNS5-brane as well as SNS1
and SD4-brane solutions of string theory by direct and double dimensional
reductions respectively. In the former case we need to use 
delocalized SM-brane 
solutions, whereas in the latter case we need to use anisotropic 
SM-brane solutions in the directions which are compactified.
\end{small}

\newpage
\sect{Introduction}

Space-like branes \cite{gs} (or S$p$-branes) are a class of time-dependent 
solutions
of string/M theory (also of some field theories) which are subject of
much interests in recent times. S$p$-branes are topological defects
localized in $(p+1)$ dimensional space-like surfaces and exist at a moment
in time. They can be understood to appear as a time-like tachyonic kink
solution of unstable D$(p+1)$-brane in string theory \cite{asen} and is 
believed to
shed light on dS/CFT correspondence \cite{as,bdm}\footnote{In three dimensions
the dS/CFT correspondence has been noted earlier in \cite{mp}.}. As for 
other time 
dependent solutions \cite{kosst}
S-branes\footnote{See \cite{lmpx} for some earlier works.}  
are also interesting 
from the cosmological point of view \cite{tw}. Various
aspects of S-branes in M/string theory have been explored in 
\cite{cgg,kmp,sr,dk}.

Just like in static case M-theory has two kinds of S-branes, namely, SM2 and 
SM5-branes and are characterized
by two parameters. On the other hand, string theory has SNS1, SNS5 and 
SD$p$-branes (as their time-like counterpart) characterized by four parameters.
These solutions were obtained in \cite{cgg} by solving the equations 
of motion of
the corresponding supergravity actions. Since type IIA string theory can be
obtained as S$^1$ reduction of M-theory, the M-theory S-branes SM2 must reduce
to SD2 of type IIA string theory under direct dimensional reduction \cite{lps}
and to
SNS1 under double dimensional reduction \cite{dhis} as in the static case. 
Similarly,
SM5-brane should reduce to SNS5-brane of type IIA string theory under direct
dimensional reduction and SD4-brane under double dimensional reduction as is
known for the static case. However, the reduction procedure can not be the 
same since the M-theory S-branes are two parameter solution, whereas the string
theory solutions are four parameter solutions \cite{cgg}. The naive 
dimensional reduction
can not produce new parameters in the reduced solutions.

The purpose of this paper is to show how starting from M-theory S-branes
one can reproduce the various string theory S-branes under dimensional 
reductions since all M-theory and string theory S-brane solutions are
known explicitly. For the BPS time-like branes, the direct dimensional 
reduction is performed along a transverse space-like isometric direction, where
isometry is produced by placing parallel branes periodically along the
to be compactified transverse direction. However, for the space-like branes
this procedure is not well-defined, so, instead we use a delocalized
SM2 and SM5-brane solutions (as has been obtained in \cite{cgg} with a 
modification) along the to be compactified transverse direction. We will 
show that the direct dimensional reduction of this solution correctly 
reproduces the known SD2\footnote{This case has also been studied in
\cite{nohta}.} and SNS5-brane solution of type IIA string theory.
The double dimensional reduction for the time-like branes is much easier to 
perform since here the compactification is done along one of the space-like
directions of the brane which is already an isometric direction. For the
space-like branes although the procedure is similar, but we will show that
we must start from anisotropic (in the to be compactified longitudinal
direction of the brane) SM-brane solutions in order to reproduce both the
SNS1 and SD4 solutions of type IIA string theory under double dimensional
reductions.

The organization of this paper is as follows. In section 2, we write the
localized S$p$-brane solutions of the gravity coupled to the dilaton and
an $n$-form field strength in $d=p+n+2$ dimensions. In $d=11,\,10$ they
represent various S-brane solutions of M/string theory in question. In
section 3, we discuss the direct dimensional reductions. The double
dimensional reductions are discussed in section 4. We conclude in section 5.

\sect{Localized S$p$-brane solutions}

In this section, we write the localized S$p$-brane solutions obtained in
\cite{cgg}
(by setting $k=q$ there) by solving the equations of motion resulting 
from $d$-dimensional
gravity coupled to the dilaton and an $n$-form field strength. The solution
in space-time dimensions $d=p+n+2$ has the symmetry ISO$(p+1)$ $\times$ 
SO$(n,1)$, ISO$(p+1)$ $\times$ ISO$(n)$ or ISO$(p+1)$ $\times$ SO$(n+1)$
depending on whether the internal spaces are hyperbolic, flat or spherical
and takes the form,
\bea
ds_d^2 &=& - e^{2A(t)} dt^2 + e^{2B(t)}(dx_1^2 + \cdots + dx_{p+1}^2) 
+ e^{2C(t)} d\Sigma_{n,\sigma}^2
\nonumber\\
\phi &=& \frac{a(d-2)}{n-1} B(t) + c_1 t + c_2\nonumber\\
F_{n} &=& b \epsilon(\Sigma_{n,\sigma}) 
\eea
In the above $A$, $B$ and $C$ are functions of time satisfying a gauge 
condition
\be
- A + (p+1) B + n C = 0
\ee
In this gauge the equations of motion simplify and the three functions
can be expressed in terms of two independent functions as,
\be
A = n g(t) - \frac{p+1}{n-1} f(t), \qquad B = f(t), \qquad C = g(t) -
\frac{p+1}{n-1} f(t)
\ee
Also in the above $d\Sigma^2_{n,\sigma}$ is the line element of the hyperbolic
($\sigma = -1$), flat ($\sigma = 0$) and spherical ($\sigma = +1$) space,
$a$ is the dilaton coupling to the $n$-form field strength, $b$ is the field
strength parameter and $\epsilon(\Sigma_{n,\sigma})$ is the unit volume element
of $\Sigma_{n,\sigma}$. By solving the equations of motion the functions
$f(t)$ and $g(t)$ are obtained as,
\bea
f(t) &=& \frac{2}{\chi} \ln \frac{\alpha}{\cosh[\frac{\chi\alpha}{2}
(t - t_0)]} + \frac{1}{\chi} \ln \frac{(d-2) \chi}{(n-1) b^2} - 
\frac{a}{\chi}(c_1t + c_2)\nonumber\\
g(t) &=& \cases{\frac{1}{(n-1)} \ln \frac{\beta}{\sinh [(n-1) \beta |t|
]}, & {\rm for} $\quad \sigma = -1$\cr
\pm \beta t, & {\rm for} $\quad \sigma = 0$\cr
\frac{1}{(n-1)} \ln \frac{\beta}{\cosh [(n-1) \beta t
]}, & {\rm for} $\quad \sigma = +1$\cr}
\eea
where $\alpha$, $\beta$, $c_1$, $c_2$, $t_0$ are integration constants. $\chi$
is defined as,
\be
\chi = 2(p+1) + \frac{a^2(d-2)}{n-1}
\ee
and the constants satisfy the following relation,
\be
\frac{p+1}{\chi} c_1^2 + \frac{(d-2) \chi \alpha^2}{2(n-1)} - n(n-1)\beta^2
=0
\ee
This is the complete localized S$p$-brane solution in $d$ space-time 
dimensions. For $d=11$, $a=c_1=c_2=0$, we get localized SM2-brane solution
for $p=2$ and $n=7$ (also $\chi=6$) and SM5-brane solution for $p=5$ and 
$n=4$ ($\chi=12$ in this case). On the other hand, for $d=10$ we get 
localized SD2 solution for $p=2$, $n=6$, $a=-1/2$, $\chi=32/5$ and SNS5
solution for $p=5$, $n=3$, $a=-1$, $\chi=16$. We also get SNS1 solution for
$p=1$, $n=7$, $a=1$, $\chi=16/3$ and SD4 solution for $p=4$, $n=4$, $a=1/2$,
$\chi=32/3$. Although we will not use the localized SM-brane solutions for
dimensional reduction as they will not produce the correct localized string
theory S-brane solutions, we will comment on them later for comparison. 
The localized string theory S-brane solutions as given in (2.1) -- (2.6)
for various values of $p$, $n$, $a$ and $\chi$ will be compared after the
dimensional reduction of appropriate M-theory S-branes. We comment that
M-theory S-branes are characterized by the parameters $\alpha$, $\beta$,
$t_0$ and $b$ with a relation between $\alpha$ and $\beta$ given by (2.6).
But $\beta$ can be eliminated from the solution by appropriate scaling of
$t$ and $x_i$ coordinates. So, the solution would depend on two parameters.
Similarly, the string theory solution can be shown \cite{sr} to depend on four
parameters $c_1$, $c_2$, $t_0$, $b$. However, in our discussion we will
keep the parameters $\alpha$ and $\beta$ explicitly for both the M-theory
and the string theory solutions.

\sect{Direct dimensional reduction}

As mentioned in the introduction in order to correctly reproduce the string 
theory S-branes we must start from a delocalized SM-brane solution and 
compactify the delocalized direction. So, instead of the form of the metric
in (2.1), we start from the metric ansatz (this is exactly the same as in
\cite{cgg}, with $q-k=1$),
\be 
ds_d^2 = -e^{2\tA(t)}dt^2 + e^{2\tB(t)}(dx_1^2+\cdots + dx_{p+1}^2)
+e^{2\tC(t)}d\Sigma^2_{n-1,\sigma} + e^{2\tD(t)}dy^2
\ee
We take $\phi=0$. The equations of motion for gravity coupled to $n$-form
field strength can be solved with the gauge condition,
\be
-\tA + (p+1)\tB + (n-1)\tC + \tD = 0
\ee
Using equations of motion and the gauge condition (3.2) the four functions
$\tA$, $\tB$, $\tC$ and $\tD$ can be expressed as,
\bea
\tA &=& (n-1)\tg(t) - \frac{p+1}{n-1}\tf(t) + n(\tc_1 t + \tc_2)\nn
\tB &=& \tf(t)\nn
\tC &=& \tg(t) - \frac{p+1}{n-1}\tf(t) + \tc_1 t + \tc_2\nn
\tD &=& -\frac{p+1}{n-1} \tf(t) + \tc_1 t + \tc_2
\eea 
Note that the delocalized solution in \cite{cgg} does not contain the linear
part $(\tc_1t+\tc_2)$ in the above functions and so, will not give the
correct string theory solution under direct dimensional reductions. The 
solution of the equations of motion give the functions $\tf(t)$ and $\tg(t)$
in the forms,
\bea
\tf(t) &=& \frac{2}{\chi} \ln \frac{\talpha}{\cosh[\frac{\chi\talpha}{2}
(t - t_0)]} + \frac{1}{\chi} \ln \frac{(d-2) \chi}{(n-1) b^2}  
\nonumber\\
\tg(t) &=& \cases{\frac{1}{(n-2)} \ln \frac{\tbeta}{\sinh [(n-2) \tbeta |t|]}
- \frac{n-1}{n-2}(\tc_1t+\tc_2), & {\rm for} $\quad \sigma = -1$\cr
\pm \tbeta t - \frac{n-1}{n-2}(\tc_1t+\tc_2), & {\rm for} $\quad \sigma = 0$\cr
\frac{1}{(n-2)} \ln \frac{\tbeta}{\cosh [(n-2) \tbeta t]}
- \frac{n-1}{n-2}(\tc_1t+\tc_2), & {\rm for} $\quad \sigma = +1$\cr}
\eea
with the parameter relation given as follows,
\be
\frac{n-1}{n-2} \tc_1^2 + \frac{(p+1)(d-2)}{n-1}\talpha^2 - (n-1)(n-2)\tbeta^2
= 0
\ee
Also, the form of the field strength is given as $F_n = b\epsilon(
\Sigma_{n-1,\sigma})\wedge dy$.

In $d=11$, eqs.(3.1) -- (3.5) represent delocalized SM2-brane solution for
$p=2$, $n=7$, $\chi=6$ and delocalized SM5-brane solution for $p=5$, $n=4$,
$\chi=12$. The 11-dimensional solution can be dimensionally reduced to 
10-dimensions along $y$-direction by the following metric relation,
\bea
ds_{11}^2 &=& e^{-2\phi/3} ds_{10,s}^2 + e^{4\phi/3} dy^2\nn
&=& e^{-\phi/6} ds_{10}^2 + e^{4\phi/3} dy^2
\eea
Where the metric $ds_{10,s}^2$ in the first line is the string frame metric
and $ds_{10}^2 = e^{-\phi/2}ds_{10,s}^2$ is the Einstein frame metric. By 
comparing (3.6) with (3.1) we obtain the dilaton as,
\be
\phi=\frac{3}{2}\left[-\frac{p+1}{n-1}\tf(t) + \tc_1t+\tc_2\right]
\ee
This would be identified with the dilaton for the 10-dimensional solutions, 
namely, the localized
SD2 and SNS5-brane solutions which we discuss separately below.

\vs{2}

\noindent{\it (a) SM2 $\rightarrow$ SD2}
\vs{2}

Note that we have $p=2$, $n=7$, $a=0$ and $\chi=6$ for SM2-brane solution and 
$p=2$, $n=6$, $a=-1/2$ and $\chi=32/5$ for SD2-brane solution. So, from (3.7)
we find,
\be
\phi = -\frac{1}{4}\ln\frac{\talpha}{\cosh[3\talpha(t-t_0)]} - \frac{1}{8}
\ln \frac{9}{b^2} + \frac{3}{2}(\tc_1t+\tc_2)
\ee
for the reduced SM2-brane solution. On the other hand, we find from (2.1)
\be
\phi = -\frac{1}{4}\ln\frac{\alpha}{\cosh[\frac{16\alpha}{5}
(t-t_0)]} - \frac{1}{8}
\ln \frac{256}{25 b^2} + \frac{15}{16}(c_1t+c_2)
\ee
for the SD2-brane solution. So, identifying (3.8) with (3.9) we get,
\be
\talpha = \frac{16}{15} \alpha, \quad \tc_1 = \frac{5}{8} c_1,
\quad \tc_2 = \frac{5}{8} c_2
\ee
We also find $\tbeta=\beta$, by examining the function $\tg(t)$
(given in (3.4)) and $g(t)$ (given in (2.4)). With this identification 
the parameter relation (3.5) becomes,
\be
\frac{6}{5}\tc_1^2 + \frac{9}{2} \talpha^2 - 30 \tbeta^2 = 0 \quad
\Rightarrow \quad \frac{15}{32} c_1^2 + \frac{128}{25} \alpha^2 - 30 \beta^2
= 0
\ee
This is exactly the parameter relation (2.6) for the SD2-brane solution. 
The form of the 10-dimensional Einstein frame metric can be calculated from
(3.6) as,
\be
ds_{10}^2 = - e^{\phi/6 + 2\tA} dt^2 + e^{\phi/6 + 2\tB}(dx_1^2+dx_2^2+dx_3^2)
+ e^{\phi/6 + 2\tC} d\Sigma_{6,\sigma}^2
\ee
Now it can be easily checked using (3.8), (3.4) and (3.3) that
\bea
\frac{1}{6}\phi + 2\tA &=& - \frac{9}{8}\tf + 12 \tg + \frac{57}{4}
(\tc_1t+\tc_2)\,\, = \,\, 12 g - \frac{6}{5}f\,\, =\,\, 2A\nn
\frac{1}{6}\phi + 2\tB &=& \frac{15}{8}\tf + \frac{1}{4}(\tc_1t+ \tc_2)
\,\,=\,\, 2f\,\, =\,\, 2B\nn
\frac{1}{6}\phi + 2\tC &=& -\frac{9}{8} \tf + 2\tg + \frac{9}{4}(\tc_1t+\tc_2)
\,\,=\,\, -\frac{6}{5}f + 2g\,\, =\,\, 2C
\eea
Where in writing the last two expressions in the above we have used eqs.(3.4),
(3.10), (2.4) and (2.3). So, the metric (3.12) takes exactly the same form
as SD2-brane metric given in (2.1). The field strength also takes the same form
as that of SD2-brane. We have thus shown how to obtain SD2-brane solution
of type IIA string theory from a delocalized SM2-brane solution of M-theory
under direct dimensional reduction.

\vs{2}

\noindent{\it (b) SM5 $\rightarrow$ SNS5}
\vs{2}

Here we note that for SM5-brane solution $p=5$, $n=4$, $a=0$ and $\chi=12$
and for SNS5-brane solution $p=5$, $n=3$, $a=-1$ and $\chi=16$. So, from (3.7)
we find that the dilaton takes the form,
\be
\phi = -\frac{1}{2}\ln\frac{\talpha}{\cosh[6\talpha(t-t_0)]} - \frac{1}{4}
\ln \frac{36}{b^2} + \frac{3}{2}(\tc_1t+\tc_2)
\ee
for the reduced SM5-brane solution. On the other hand, we find from (2.1)
\be
\phi = -\frac{1}{2}\ln\frac{\alpha}{\cosh[8\alpha
(t-t_0)]} - \frac{1}{4}
\ln \frac{64}{b^2} + \frac{3}{4}(c_1t+c_2)
\ee
for the SNS5-brane solution. So, identifying (3.14) with (3.15) we get,
\be
\talpha = \frac{4}{3} \alpha, \quad \tc_1 = \frac{1}{2} c_1,
\quad \tc_2 = \frac{1}{2} c_2
\ee
Also, as before we have $\tbeta=\beta$.
With this identification the parameter relation for the delocalized SM5-brane
solution (3.5) takes the following form using (3.16),
\be
\frac{3}{2}\tc_1^2 + 18 \talpha^2 - 6 \tbeta^2 = 0 \quad
\Rightarrow \quad \frac{3}{8} c_1^2 + 32 \alpha^2 - 6 \beta^2
= 0
\ee
This is exactly the parameter relation of SNS5-brane solution given in (2.6). 
The 10-dimensional Einstein frame metric can be obtained from (3.6) as,
\be
ds_{10}^2 = - e^{\phi/6 + 2\tA} dt^2 + e^{\phi/6 + 2\tB}(dx_1^2+\cdots +dx_6^2)
+ e^{\phi/6 + 2\tC} d\Sigma_{3,\sigma}^2
\ee
It can be checked using (3.14), (3.4) and (3.3) that
\bea
\frac{1}{6}\phi + 2\tA &=& - \frac{9}{2}\tf + 6 \tg + \frac{33}{4}
(\tc_1t+\tc_2)\,\, = \,\, 6 g - 6 f\,\, =\,\, 2A\nn
\frac{1}{6}\phi + 2\tB &=& \frac{3}{2}\tf + \frac{1}{4}(\tc_1t+ \tc_2)
\,\,=\,\, 2f\,\, =\,\, 2B\nn
\frac{1}{6}\phi + 2\tC &=& -\frac{9}{2} \tf + 2\tg + \frac{9}{4}(\tc_1t+\tc_2)
\,\,=\,\, -6 f + 2g\,\, =\,\, 2C
\eea
Where in writing the last two expressions in the above we have used eqs.(3.4),
(3.16), (2.4) and (2.3). So, the metric (3.18) takes exactly the same form
as SNS5-brane metric given in (2.1). The field strength also takes the same 
form as that of SNS5-brane solution. So, we have obtained SNS5-brane solution
of type IIA string theory starting form the delocalized SM5-brane solution
of M-theory by direct dimensional reduction.

\sect{Double dimensional reduction}

In the case of double dimensional reduction we compactify the M-theory 
along one
of the space-like directions of the brane. But, it is clear that if we start 
from the isotropic M-brane solution given in (2.1), then the dilaton (3.6)
in the reduced theory will not contain the linear time-dependent part
(of the form $c_1t+c_2$) and we will not get the right form of the dilaton
(2.1) of the string theory S-brane. However, we show that if we use the
anisotropic (along the to be compactified direction) M-theory S-brane 
solution\footnote{This has also been recognized for $\sigma=-1$ in different
supergravity S-brane solutions discussed in \cite{kmp}.},
then we correctly reproduce the string theory S-brane under double dimensional
reduction.

So, instead of (2.1), we take the metric ansatz of the M-theory S-brane
solution as,
\be 
ds_d^2 = -e^{2\hA(t)}dt^2 + e^{2\hB(t)}(dx_1^2+\cdots + dx_p^2)
+e^{2\hC(t)}d\Sigma^2_{n,\sigma} + e^{2\hD(t)}dy^2
\ee
where we identify $x_{p+1} \equiv y$ as the to be compactified direction.
Here also, we take $\phi=0$. The equations 
of motion is the same as in eqs.(3) -- (5) of ref.\cite{cgg} with $\phi=a=0$.
However, the non-vanishing components of the Ricci tensor for the above
metric takes the form
\bea
R_{tt} &=& -p(\ddot{\hB}+\dot{\hB}^{\,2}-\dot{\hA}\dot{\hB})-n(\ddot{\hC}+
\dot{\hC}^2-\dot{\hA}\dot{\hC})-(\ddot{\hD}+\dot{\hD}^2-\dot{\hA}\dot{\hD})\nn
R_{xx} &=& e^{2\hB-2\hA}\left[\ddot{\hB}-\dot{\hA}\dot{\hB}+p\dot{\hB}^2
+n\dot{\hB}\dot{\hC}+\dot{\hB}\dot{\hD}\right]\nn
R_{yy} &=& e^{2\hD-2\hA}\left[\ddot{\hD}-\dot{\hA}\dot{\hD}+p\dot{\hB}\dot{\hD}
+n\dot{\hD}\dot{\hC}+\dot{\hD}^2\right]\nn
R_{ab} &=& \left\{e^{2\hC-2\hA}\left[\ddot{\hC}-\dot{\hA}\dot{\hC}+
p\dot{\hB}\dot{\hC}
+n\dot{\hC}^2+\dot{\hC}\dot{\hD}\right] + \sigma(n-1)\right\}\bar{g}_{ab}
\eea
where $\bar{g}_{ab}$ is the metric of the hyperspace $\Sigma_{n,\sigma}$
and the Ricci tensor for this space is given by, $\bar{R}_{ab} = 
\sigma(n-1)\bar{g}_{ab}$. The equations of motion simplifies under the gauge 
condition, 
\be
-\hA + p \hB + n \hC + \hD = 0
\ee
Using this and the equations of motion, the above functions can be expressed
as,
\bea
\hA &=& n\hg(t) - \frac{p+1}{n-1}\hf(t) + (n+1)(\hc_1 t + \hc_2)\nn
\hB &=& \hf(t)\nn
\hC &=& \hg(t) - \frac{p+1}{n-1}\hf(t) + \hc_1 t + \hc_2\nn
\hD &=& \hf(t) + \hc_1 t + \hc_2
\eea 
By solving the equations of motion we obtain,
\bea
\hf(t) &=& \frac{2}{\chi} \ln \frac{\halpha}{\cosh[\frac{\chi\halpha}{2}
(t - t_0)]} + \frac{1}{\chi} \ln \frac{(d-2)\chi}{(n-1) b^2}  
- \frac{1}{p+1}(\hc_1t+\hc_2)\nonumber\\
\hg(t) &=& \cases{\frac{1}{(n-1)} \ln \frac{\hbeta}{\sinh [(n-1) \hbeta |t|]}
- \frac{n}{n-1}(\hc_1t+\hc_2), & {\rm for} $\quad \sigma = -1$\cr
\pm \hbeta t - \frac{n}{n-1}(\hc_1t+\hc_2), & {\rm for} $\quad \sigma = 0$\cr
\frac{1}{(n-1)} \ln \frac{\hbeta}{\cosh [(n-1) \hbeta t]}
- \frac{n}{n-1}(\hc_1t+\hc_2), & {\rm for} $\quad \sigma = +1$\cr}
\eea
where the parameters satisfy,
\be
\frac{p}{p+1} \hc_1^2 + \frac{(p+1)(d-2)}{n-1}\halpha^2 - n(n-1)\hbeta^2
= 0
\ee
The field strength is given as $F_n = b \epsilon(\Sigma_{n,\sigma})$. In
$d=11$, eqs.(4.1) -- (4.6) represent the anisotropic SM2-brane solution
for $p=2$, $n=7$, $\chi=6$ and anisotropic SM5-brane solution for $p=5$,
$n=4$, $\chi=12$. The dimensional reduction is performed along the brane
direction $y$ and the relation between the 11-dimensional metric and the
10-dimensional metric is as given in (3.6). By comparing (3.6) with (4.1)
we obtain the dilaton as,
\be
\phi = \frac{3}{2} \hf + \frac{3}{2}(\hc_1t + \hc_2)
\ee
This would be identified with the dilaton for the 10-dimensional solution 
i.e. the localized SNS1
and SD4-brane solution and we discuss the two cases separately below.

\vs{2}

\noindent{\it (a) SM2 $\rightarrow$ SNS1}

\vs{2}

In this case we have $p=2$, $n=7$, $a=0$, $\chi=6$ for SM2-brane solution and
$p=1$, $n=7$, $a=1$, $\chi=16/3$ for SNS1-brane solution (given in (2.1)).
So, from (4.7) and (4.5) we obtain
\be
\phi = \frac{1}{2}\ln\frac{\halpha}{\cosh[3\halpha(t-t_0)]} + \frac{1}{4}
\ln \frac{9}{b^2} + (\hc_1t+\hc_2)
\ee
for the dimensionally reduced anisotropic SM2-brane. On the other hand, 
from (2.1) we find,
\be
\phi = \frac{1}{2}\ln\frac{\alpha}{\cosh[\frac{8\alpha}{3}
(t-t_0)]} + \frac{1}{4}
\ln \frac{64}{9 b^2} + \frac{3}{4}(c_1t+c_2)
\ee
for the SNS1-brane. Identifying (4.8) with (4.9) we find,
\be
\halpha = \frac{8}{9} \alpha, \quad \hc_1 = \frac{3}{4} c_1,
\quad \hc_2 = \frac{3}{4} c_2
\ee
and comparing the function $\hg(t)$ in (4.5) and $g(t)$ in (2.1)
we get $\hbeta=\beta$. Now using (4.10) the parameter relation (4.6) reduces to,
\be
\frac{2}{3}\hc_1^2 + \frac{9}{2} \halpha^2 - 42 \hbeta^2 = 0 \quad
\Rightarrow \quad \frac{3}{8} c_1^2 + \frac{32}{9} \alpha^2 - 42 \beta^2
= 0
\ee
This is exactly the parameter relation of SNS1-brane solution given in (2.6). 
The 10-dimensional Einstein frame metric (3.6) takes the form,
\be
ds_{10}^2 = - e^{\phi/6 + 2\hA} dt^2 + e^{\phi/6 + 2\hB}(dx_1^2+dx_2^2)
+ e^{\phi/6 + 2\hC} d\Sigma_{7,\sigma}^2
\ee
Now using (4.7) and (4.4) we find,
\bea
\frac{1}{6}\phi + 2\hA &=& - \frac{3}{4}\hf + 14 \hg + \frac{65}{4}
(\hc_1t+\hc_2)\,\, = \,\, - \frac{2}{3} f + 14g\,\, =\,\, 2A\nn
\frac{1}{6}\phi + 2\hB &=& \frac{9}{4}\hf + \frac{1}{4}(\hc_1t+ \hc_2)
\,\,=\,\, 2f\,\, =\,\, 2B\nn
\frac{1}{6}\phi + 2\hC &=& -\frac{3}{4} \hf + 2\hg + \frac{9}{4}(\hc_1t+\hc_2)
\,\,=\,\, -\frac{2}{3} f + 2g\,\, =\,\, 2C
\eea
Where in writing the last two expressions above we have used eqs.(4.5),
(4.10), (2.4) and (2.3). We therefore find that the metric in (4.12) matches
exactly with that of SNS1-brane solution given in (2.1). The 7-form field
strength also matches trivially. This therefore shows that the double 
dimensional reduction of an anisotropic SM2-brane solution indeed correctly 
reproduces the SNS1-brane solution of type IIA string theory.

\vs{2}

\noindent{\it (b) SM5 $\rightarrow$ SD4}

\vs{2}

Here also we employ the same procedure as in the previous subsection. In this
case $p=5$, $n=4$, $a=0$, $\chi=12$ for SM2-brane solution and $p=4$, $n=4$, 
$a=1/2$, $\chi=32/3$ for SD4-brane solution (given in (2.1)). The dilaton 
for the dimensionally reduced anisotropic SM5-brane can be obtained from (4.7)
and (4.5) as,
\be
\phi = \frac{1}{4}\ln\frac{\halpha}{\cosh[6\halpha(t-t_0)]} + \frac{1}{8}
\ln \frac{36}{b^2} + \frac{5}{4}(\hc_1t+\hc_2)
\ee
The form of the dilaton for the SD4-brane solution can be obtained
from (2.1) as,
\be
\phi = \frac{1}{4}\ln\frac{\alpha}{\cosh[\frac{16\alpha}{3}
(t-t_0)]} + \frac{1}{8}
\ln \frac{256}{9 b^2} + \frac{15}{16}(c_1t+c_2)
\ee
Identifying these two we obtain,
\be
\halpha = \frac{8}{9} \alpha, \quad \hc_1 = \frac{3}{4} c_1,
\quad \hc_2 = \frac{3}{4} c_2
\ee
Also as before we get $\hbeta=\beta$.
So, using (4.16) the parameter relation for SM5-brane solution (4.6) gives,
\be
\frac{5}{6}\hc_1^2 + 18 \halpha^2 - 12 \hbeta^2 = 0 \quad
\Rightarrow \quad \frac{15}{32} c_1^2 + \frac{128}{9} \alpha^2 - 12 \beta^2
= 0
\ee
We note that this is precisely the parameter relation for the SD4-brane 
solution as can be seen from (2.6). Now we can check whether the dilaton as 
well as the parameter identification correctly reproduces the 10-dimensional
SD4-brane metric from (3.6). 
The Einstein frame metric in (3.6) has the form,
\be
ds_{10}^2 = - e^{\phi/6 + 2\hA} dt^2 + e^{\phi/6 + 2\hB}(dx_1^2+\cdots+dx_5^2)
+ e^{\phi/6 + 2\hC} d\Sigma_{4,\sigma}^2
\ee
It can be easily checked using (4.7) and (4.16) that
\bea
\frac{1}{6}\phi + 2\hA &=& - \frac{15}{4}\hf + 8 \hg + \frac{41}{4}
(\hc_1t+\hc_2)\,\, = \,\, - \frac{10}{3} f + 8g\,\, =\,\, 2A\nn
\frac{1}{6}\phi + 2\hB &=& \frac{9}{4}\hf + \frac{1}{4}(\hc_1t+ \hc_2)
\,\,=\,\, 2f\,\, =\,\, 2B\nn
\frac{1}{6}\phi + 2\hC &=& -\frac{15}{4} \hf + 2\hg + \frac{9}{4}(\hc_1t+\hc_2)
\,\,=\,\, -\frac{10}{3} f + 2g\,\, =\,\, 2C
\eea
In the last two expressions above we have used eqs.(4.5),
(4.16), (2.4) and (2.3). This shows that we indeed obtain SD4 metric, dilaton,
and the 4-form field strength (this matches trivially) starting from the
anisotropic SM5-brane solution by double dimensional reduction.
 
\sect{Conclusion}

In this paper we have studied the direct as well as the double dimensional
reduction of the M-theory S-branes to string theory S-branes. This procedure
is well-known for the usual time-like branes, but we pointed out that the
reduction procedure is quite different for the S-branes. The difference can be
understood since both M-theory and string theory S-brane solutions are 
known explicitly and it is also known that M-theory solutions are characterized
by two parameters whereas the string theory solutions are characterized by
four parameters. Although the physical meaning of these parameters is not
well understood, it is clear that new parameters can not be produced by only 
dimensional reduction. New parameters can be introduced if instead of
localized SM-brane solutions given in eq.(2.1), we start from delocalized
(in one of the transverse to be compactified space-like directions of the
brane) or anisotropic (in one of the longitudinal to be compactified 
directions of the brane) solutions. Inspection of the equations of motion
suggests that the delocalization or the anisotropization (along one direction
which is to be compactified) of the SM-branes produces  exactly the required 
number of parameters for the string theory S-branes. This is exactly what
we have done and thus we have shown that these solutions i.e. SM2 and SM5
correctly reproduces the string theory S-branes namely, SD2 and SNS5-brane
under direct dimensional reduction and SNS1 and SD4-brane under double 
dimensional reduction. For direct dimensional reduction we have used the
delocalized solution and for double dimensional reduction we have used the 
anisotropic solution.

It should be noted that since the dilaton of the string theory is related to
the radius of the compactified eleventh dimension by $e^{2\phi/3} \sim
R_{11}$ and it is time dependent, it is quite crucial to see whether the
compactification is achieved by looking at whether $e^{\phi} \sim R_{11}^{3/2}
\ll 1$. As $t$ varies, we notice from (3.7) and (4.7) that it would depend on
the parameters $\alpha$ and $c_1$. It is conceivable that there might exist
some range of $t$, where $R_{11}$ remains large and we will see the full
11-dimensional theory instead of the 10-dimensional one. However, in 11
dimensions the decompactified (or `uplifted') string theory S-branes 
would not be the
localized, isotropic SM-branes, but the delocalized or anisotropic branes.

\section*{Acknowledgements}

I would like to thank Somdatta Bhattacharya and Sudipta Mukherji for 
discussions.

\end{document}